\documentclass[12pt]{article}
\usepackage[colorlinks,linkcolor=blue,citecolor=blue,urlcolor=blue,bookmarks,bookmarksnumbered]{hyperref}
\usepackage[paper=letterpaper,margin=1.0in]{geometry}
\usepackage{graphicx,xcolor}

\usepackage{amsfonts}
\usepackage{amsmath,amssymb,amsthm,bm,multirow,longtable}
\usepackage{epsfig}
\usepackage{epstopdf}
\usepackage{color}
\usepackage{cite}

\newcommand{\be}{\begin{equation}}
\newcommand{\ee}{\end{equation}}
\newcommand{\bea}{\begin{eqnarray}}
\newcommand{\eea}{\end{eqnarray}}
\newcommand{\ba}{\begin{array}}
\newcommand{\ea}{\end{array}}
\newcommand{\ben}{\begin{enumerate}}
\newcommand{\een}{\end{enumerate}}
\newcommand{\bi}{\begin{itemize}}
\newcommand{\ei}{\end{itemize}}
\newcommand{\bc}{\begin{center}}
\newcommand{\bfig}{\begin{figure}}
\newcommand{\efig}{\end{figure}}
\newcommand{\bq}{\begin{quotation}}
\newcommand{\eq}{\end{quotation}}
\newcommand{\bt}{\begin{table}}
\newcommand{\et}{\end{table}}
\newcommand{\btab}{\begin{tabular}}
\newcommand{\etab}{\end{tabular}}
\newcommand{\bs}{\begin{slide}}
\newcommand{\es}{\end{slide}}

\newcommand{\pa}{\partial}

\newcommand{\IR}{\mathbb{R}}

\newcommand{\X}{\mathbb{X}}

\def\pa{\partial}

\newcommand{\rd}{\mathrm{d}}

\def\s{\sigma}

\let\a=\alpha

\let\ba=\overline

\def\rd{{\rm d}}
\def\define{\buildrel{\smash{\scriptscriptstyle\rm def}}\over=}

\let\F=\Phi
\def\e{\epsilon}
\let\F=\Phi

\def\inv#1{{\textstyle{1\over#1}}}

\let\l=\lambda
\let\L=\Lambda

\let\q=\theta

\let\t=\tau
\let\vd=\partial

\def\hx{\mathord{\hat x}}
\def\tx{\mathord{\tilde x}}
\def\htx{\mathord{\hat{\tilde x}}}

\let\w=\omega

\def\IR{\relax\leavevmode{\rm I\kern-.18em R}}
\def\ZZ{\relax\leavevmode
       \ifmmode\mathchoice
       {\hbox{\sf Z\kern-.4em Z}}
       {\hbox{\sf Z\kern-.4em Z}}
       {\lower.9pt\hbox{\scriptsize\sf Z\kern-.36em Z}}
       {\lower1.2pt\hbox{\tiny\sf Z\kern-.36em Z}}
       \else{\sf Z\kern-.4em Z}\fi}
\def\RR{\relax\leavevmode
       \ifmmode\mathchoice
       {\hbox{\sf R\kern-.4em R}}
       {\hbox{\sf R\kern-.4em R}}
       {\lower.9pt\hbox{\scriptsize\sf R\kern-.36em R}}
       {\lower1.2pt\hbox{\tiny\sf R\kern-.36em R}}
       \else{\sf R\kern-.4em R}\fi}

\def\resetby#1#2{\@addtoreset{#2}{#1}}
\def\seceq{\@addtoreset{equation}{section}
              \def\theequation{\thesection.\arabic{equation}}}

\def\Label#1{\label{#1}%
                \smash{\hbox to0pt{\raise1ex\hbox{\tiny[#1]}\hss}}}
\def\noLabels{\let\Label=\label}

\def\TeV{\text{T\kern0pte\kern-1ptV}}

\begin{document}
\thispagestyle{empty}

\vskip 0.5cm
\title{\bf Dark Energy and String Theory}
\author{\normalsize
        Per Berglund\textsuperscript1,
        Tristan H{\"u}bsch\textsuperscript2 and
        Djordje Mini{\'c}\textsuperscript3\\*
        \footnotesize\textsuperscript1\it\,Department of Physics and Astronomy, University of New Hampshire, Durham, NH 03824, U.S.A.\\*[-5pt]
        \footnotesize\textsuperscript2\it\,Department of Physics and Astronomy, Howard University, Washington, D.C.  20059, U.S.A.\\*[-5pt]
        \footnotesize\textsuperscript3\it\,Department of Physics, Virginia Tech, Blacksburg, VA 24061, U.S.A.
} 
\date{\small\today}

\maketitle
\begin{abstract}\noindent
A radiatively stable de~Sitter spacetime is constructed by considering 
an intrinsically non-commutative and generalized-geometric formulation of string theory, 
which is related to a family of F-theory models
endowed with non-trivial anisotropic axion-dilaton backgrounds. In particular, the curvature of the canonically conjugate dual space provides for a positive cosmological constant to leading order, that satisfies a radiatively stable see-saw-like formula, which in turn induces the dark energy in the observed spacetime. 
We also comment on the non-commutative phase of the non-perturbative formulations of string theory/quantum gravity 
implied by this approach.
\end{abstract}

\paragraph{Introduction:} String theory~\cite{Polchinski:1998rq} still represents one of the most promising approaches for a consistent theory of
quantum gravity and matter. Yet, ever since the seminal discovery of dark
energy in the late 1990s~\cite{Riess:1998cb}, string theory has been attempting to deal
with this central ingredient of fundamental physics.
(For the most recent measurements of the Hubble constant and the associated discrepancies (see~\cite{Riess:2016jrr}). 
The existence of de~Sitter space as a solution in string theory (and dark energy in the observable universe) is still considered an 
outstanding open question~\cite{Danielsson:2018ztv}, and the interest in this fundamental issue has been recently reignited in~\cite{Obied:2018sgi} (see also~\cite{Andriot:2019wrs}). 
In this letter we argue that  
a generic, non-commutative generalized geometric phase-space formulation of string theory,
leads to the low energy effective action valid at long distances of the observed accelerated universe (focusing on the relevant $3{+}1$-dimensional 
spacetime $X$, of the ${+}\,{-}\,{-}\,{-}$ signature):
\be
S_{\text{eff}} =\frac{-1}{8 \pi G}\int_X \sqrt{-g}
  \big( \L + {\textstyle\frac{1}2} R + {\cal O}(R^2)
\big),
 \label{e:Seff}
\ee
with $\L$ the positive cosmological constant and
the ${\cal O}(R^2)$ correction terms arising from string theory~\cite{rF79b}.
In particular, this new approach is realized in certain stringy-cosmic-string-like~\cite{rGSVY} toy models,
which can be viewed as illustrative of a generic non-commutative phase of F-theory~\cite{rBHM7}.

\paragraph{Generic formulation of string theory:} The generalized geometric formulation of string theory we have in mind has been recently discussed 
in~\cite{Freidel:2013zga, Freidel:2015uug, Freidel:2016pls, 
Freidel:2017xsi, Freidel:2017wst}, 
and derives from  a 
{\it chiral} world-sheet description
\be
S_{\text{str}}\,{=}\,\frac{1}{4\pi}\int_{\Sigma}
 \Big[\pa_{\tau}{\X}^{A} (\eta_{AB}+\w_{AB})\pa_\s\X^B
- \pa_\s\X^A\,H_{\!AB}\,\pa_\s\X^B\Big], 
\label{e:MSA}
\ee 
where $\Sigma$ is the worldsheet, and $\X^A$, ($A=1,...,26$, for the critical bosonic string) combine the sum ($x^a$) and the difference ($\tx_a$) of the left- and right-movers on the string. 
The mutually compatible dynamical fields $\w_{AB}(\X),\eta_{AB}(\X)$ and $H_{AB}(\X)$ are:
the antisymmetric symplectic structure $\w_{AB}$,
the symmetric polarization metric $\eta_{AB}$ and
the doubled symmetric metric $H_{\!AB}$, respectively, defining the so-called Born geometry~\cite{Freidel:2013zga}.
Quantization renders the doubled ``phase-space'' operators $\hat{\X}^A=(\hx^a/\l, \htx_b/\l)$ inherently non-commutative, inducing~\cite{Freidel:2017wst}:
\be
[ \hat{\X}^A, \hat{\X}^B] = i \w^{AB},
\label{e:CnCR}
\ee
or, in components, for constant non-zero $\w^{AB}$,
\bea
[\hx^a,\htx_b]=2\pi i\l^2 \delta^a_b,\quad
[\hx^a,\hx^b]=0=[\htx_a,\htx_b],
\label{e:CnCR1}
\eea
where $\l$ denotes the fundamental length scale, such as the Planck scale,
so that $\e=1/\l$ is the corresponding fundamental energy scale
and the string tension is 
$\a' = \l/{\e}=\lambda^2$.
This 
fully spacetime covariant formulation was
found by examining the simplest example of the canonical free string compactified on a circle, in an intrinsically T-duality covariant formulation of the Polyakov string 
and was independently confirmed by examining the algebra of vertex operators in the 2d CFT of a free string compactified on a circle~\cite{
Freidel:2013zga, Freidel:2015uug, Freidel:2016pls, 
 Freidel:2017xsi, Freidel:2017wst}.
In this formulation
all effective fields must be regarded a priori as 
{\em\/bi-local\/} $\phi(x, \tx)$~\cite{Freidel:2016pls}, subject to~\eqref{e:CnCR1}, and therefore inherently non-local in the conventional $x^a$-spacetime. Such non-commutative field theories~\cite{Douglas:2001ba, Grosse:2004yu} generically display a mixing between the ultraviolet (UV) and infrared (IR) physics
with continuum limits defined via a double-scale renormalization group (RG) and the self-dual fixed points~\cite{Grosse:2004yu,Freidel:2017xsi}. 

\paragraph{Generic string theory and dark energy:} We  
now argue that the generalized geometric formulation of string theory discussed above provides for an effective description of dark energy that is consistent with a de~Sitter spacetime. This is due to the theory's chirally and non-commutatively~\eqref{e:CnCR1} doubled realization of the target space 
and the stringy effective action on the doubled non-commutative~\eqref{e:CnCR1} spacetime $(x^a,\tx^a)$ 
\be
  S_{\text{eff}}^{\textit{nc}}
  =\iint \text{Tr} \sqrt{g(x,\tx)}\, \big[R(x,\tx) +L_m(x,\tx) +\dots\big],
\label{e:ncEH}
\ee
where the ellipses denote higher-order curvature terms induced by string theory~\eqref{e:Seff}.  (Here we have included the matter Lagrangian $L_m$
as well.) Owing to~\eqref{e:CnCR1}, this $S_{\text{eff}}^{nc}$ clearly expands into numerous terms with different powers of $\lambda$, which upon $\tx$-integration and from the $x$-space vantage point produce
various effective terms.
To lowest ({\em\/zeroth\/}) order of the expansion in the non-commutative parameter $\lambda$
of $S_{\text{eff}}^{\textit{nc}}$ takes the form:
\be
S_d = - \iint\! \sqrt{-g(x)} \sqrt{-\tilde{g}(\tx)} [R(x) + \tilde{R}(\tx)],
\label{e:TsSd}
\ee
a result which was first obtained almost three decades ago,  effectively neglecting $\w_{AB}$ in~\eqref{e:CnCR} 
by assuming that $[\hat x^a,\htx]=0$~\cite{Tseytlin:1990nb,
Tseytlin:1990hn}. In this leading limit, the $\tx$-integration in the first term of~\eqref{e:TsSd} defines the gravitational constant $G_N$, and in the second term produces a {\it positive} cosmological constant constant $\L>0$. It also follows that the weakness of gravity is determined by the size of the canonically conjugate dual space, while the smallness of the cosmological constant is given by its curvature. 
(Higher order terms in $\lambda$ produce various forms of dark energy~\cite{Joyce:2014kja}
and this may even provide for a way of addressing the recent conflicting measurements of the Hubble constant~\cite{Riess:2016jrr}.)
Given this action, we may proceed reinterpreting Tseytlin's work~\cite{Tseytlin:1990hn}: integrate out the dual spacetime coordinates, write
the effective action as
  $\bar{S} \sim \tilde{V} \int_X\! \sqrt{-g(x)} R(x)+...,$
where
  $\tilde{V} = \int_{\tilde X}\! \sqrt{-\tilde{g}(\tx)},$
and then relate the dual spacetime volume to the observed spacetime volume as
$\tilde{V} \sim V^{-1}$. This produces 
an ``intensive'' effective action~\cite{Tseytlin:1990hn}
\be
 \bar{S} 
= \frac{\int_X\! \sqrt{-g(x)} \big(R(x) +L_m(x)\big)}{ \int_X\! \sqrt{-g(x)}}+\dots
 \label{e:TsLb}
\ee
see also~\cite{Lombriser:2019jia}.
 By concentrating on the classical description first (we discuss below 
 quantum corrections and the central role of intrinsic non-commutativity in string theory) we get the following Einstein equations~\cite{Tseytlin:1990hn}
\begin{equation}
 R_{ab} - \frac{1}{2} R g_{ab} +T_{ab} + \frac{1}{2} \bar{S}\, g_{ab} =0,\qquad
T_{ab} \overset{\scriptscriptstyle\text{def}}= \frac{\partial L_m}{\partial g^{ab}} - \frac{1}{2} L_m\,g_{ab}.
\end{equation}
In what follows we assume that $L_m$ contains, apart from the
Standard Model-like matter, also the hallmark stringy matter --- the axion-dilaton system.
We emphasize that our reinterpretation of~\cite{Tseytlin:1990hn}
does not follow the original presentation and intention.
In particular, we directly relate the intensive action~\eqref{e:TsLb} to the cosmological constant, 
$\bar{S}\sim \Lambda$.

\paragraph{A specific string model:} This result turns out to be directly related to the prediction of a class of a particular 
discretuum of toy models~\cite{rBHM1, rBHM5, rBHM6, rBHM7} that aim to realize de~Sitter space in string theory,
and which naturally capture several of the features of the above non-commutatively generalized phase-space reformulation of string theory. 
This family of models is constructed by starting with an
  F-theoretic~\cite{rFTh} type-IIB string theory spacetime, $W^{3,1} \times Y^4\times Y_\bot^2(\times T^2)$, where the complex structure of the zero-size ``hidden'' $T^2$ fiber of F-theory is identified with the axion-dilaton $\t\define\a\,{+}\,ie^{-\F}$ modulus.
Specifically, we compactify on $Y^4=\text{K3}$ or $T^4$ and let 
 the observable spacetime $W^{3,1}$ (via warped metric) vary over $Y_\bot^2$,  and $Y_\bot^2\to S^1 \times Z$, with the polar parametrization $re^{i\q}=\ell e^{z+i\q}$,
while $Y^4$ preserves supersymmetry. Finally, we deform $\t$ to vary {\em\/non-holomorphically,} only over $S^1\,{\subset}\,Y^2$.
By cross-patching two distinct solutions and by deforming the apparently singular metric into de~Sitter space, we get the final non-supersymmetric solution.
The codimension-2 solution $W^{3,1}\rtimes(S^1\times Z)$, has a positive cosmological constant, $\L$, along $W^{3,1}$, and the warped metric is~\cite{Cohen:1999ia}
\begin{equation}
 \rd s^2 = A^2(z)\, \bar g_{ab}\,\rd x^a \rd x^b - \ell^2 B^2(z)\,(\rd z^2 + \rd\q^2),
\end{equation}
where the metric on $W^{3,1}$ reads $\bar g_{ab}\,\rd x^a \rd x^b = \rd x_0^2 - e^{2\sqrt{\L}\,x_0}\,(\rd x_1^2 +\rd x_{2}^2 + \rd x_{3}^2) $,
and where $z=\log(r/\ell) \in Z$.
The two explicit solutions for $\t$ are~\cite{rBHM1}
\begin{alignat}9
 \t_I(\q) &=b_0+i\,g_s^{-1}\,e^{\w(\q-\q_0)}, \quad\text{and} \label{e:tau1}\\
\t_{I\!I}(\q) 
&=\big(b_0\pm g_s^{-1}\tanh[\w(\q{-}\q_0)]\big) \pm i\,g_s^{-1}(\cosh[\w(\q{-}\q_0)])^{-1}. \label{e:tau2}
\end{alignat}
Given the stringy SL$(2;\mathbb{Z})$ monodromy of the axion-dilaton system over a transversal 2-plane $Y_\bot^2$ in the spacetime, these toy models exhibit S-duality.
 In generalizations where various moduli fields replace the axion-dilaton system,
this directly implies T-duality, 
which is covariantly realized in the generic phase-space approach~\eqref{e:MSA}--\eqref{e:CnCR1}.

We emphasize that these models are
a deformation of the stringy cosmic string, and as such represent effective
stringy solutions (in the sense of~\cite{Polchinski:1991ax}) and not just IIB supergravity solutions. That is, our solutions are
indeed found as deformations of certain classic F-theory backgrounds, but as codimension-2 solutions
they can be viewed as effective stringy solutions with an effective ``world-sheet''
description that is, to lowest order,  
doubled and generically non-commutative (as described by
equation~\eqref{e:MSA}). Thus our deformed stringy cosmic string solutions
are naturally equipped with a generalized geometric (and
non-commutatively doubled) spacetime structure, which to lowest order of the
doubled target space description directly connects to~\cite{Tseytlin:1990hn}.
Therefore certain generic features of this doubled description, such as the intensive effective
action, directly translate into certain geometric features of our models, discussed below.

In these string models 
the cosmological constant within the codimension-2 brane-world is determined by the anisotropy $\Delta \omega$ of the axion-dilaton system 
whose effective energy momentum tensor is given via
\begin{equation}
 \widetilde{T}_{\mu\nu}\overset{\scriptscriptstyle\text{def}}=
 T_{\mu\nu}{-}{\textstyle\frac12}g_{\mu\nu}\,g^{\rho\s}T_{\rho\s}
 ={\cal G}_{\t\bar{\t}}\,\vd_{\mu}\t\vd_{\nu}\bar\t
 =\hbox{diag}[0,\cdots,\,0,\,\inv4\w^2\ell^{-2}],
\end{equation}
with ${\cal G}_{\t \bar{\t}}=-1/(\t{-}\bar\t)^2$, and
where $\ell$ is  the characteristic length-scale in the transversal 2-plane $Y_\bot^2$~\cite{rBHM5,rBHM6,rBHM7}:
\be
  \Lambda \sim \frac {{\Delta \omega}^2}{\ell^2}~~\text{implies}~~
  M_\L\,{\sim}\,M^2/M_P,
\label{e:seesaw}
\ee
relating the mass scales of the vacuum energy/cosmological constant ($M_\L$), particle physics, i.e., Standard Model ($M$), and the Planck scale ($M_P$). This {\it see-saw} formula can be seen to arise in two ways:
First, the formula~\eqref{e:seesaw}  may be understood as a consequence of dimensional transmutation, whereby the (modified) logarithmic nature of the transversal Green's function~\cite{rBHM1} (characteristic only of codimension-2 solutions) relates the length-scales $\ell$ and $\sqrt{\Lambda}$~\cite{rBHM5}.
Alternatively, the see-saw formula~\eqref{e:seesaw} follows from adapting Tseytlin's result for $\bar{S}$ to the models of~\cite{rBHM5,rBHM6,rBHM7}: In the denominator of~\eqref{e:TsLb}, the volume of the transversal 2-plane produces the length scale $\int_{Y_\bot^2}\!\sqrt{-g(x)}\,{\propto}\,\ell^2$; the numerator 
(with ${\Delta \omega}^2 \define \big(\omega^2 - \omega^2_c\,A^2{(z\,{=}\,0)}\big) $)
\be
  \int_{Y_\bot^2}\! \sqrt{-g(x)} 
  \big(R(x) +L_m\big) \propto {\Delta \omega}^2,
\ee
reproduces the anisotropy variance of these axion-dilaton profiles, whereas the remaining volume-integration renormalizes the Newton constant as required in~\cite{rBHM1,rBHM5}.
The anisotropy $\omega$ determines the above axion-dilaton stress tensor for the de~Sitter solution, and asymptotes to the Minkowski
cosmic brane limit $\omega_{c}$ at $z \to 0$. 
Note that in the F-theory limit, 
$\omega \to 0$ and $\omega_c \to 0$.
This singular supersymmetric configuration is deformed into a de~Sitter background by turning on an anisotropic axion-dilaton profile~\eqref{e:tau1}--\eqref{e:tau2}.
Thus $\Lambda$ that figures in the see-saw formula can be understood as being related to the cosmological breaking of supersymmetry.  
We stress that our discussion gives an argument for the existence of
de~Sitter background in string theory, albeit in its generic generalized-geometric and intrinsically
non-commutative formulation, which from the effective spacetime description is described
by our stringy models. One of the features of this doubled and generalized geometric 
description is that the effective spacetime action is intensive (as opposed to extensive), 
which directly translates into the see-saw formula for the cosmological constant~\eqref{e:seesaw}.

\paragraph{Radiative stability and beyond:} 
These results from the commutative limit are not stable under loop corrections, which has been addressed in the recent work of Kaloper and Padilla (called the sequester mechanism) who also extended these results to loops of arbitrary order, in the effective field theory~\cite{Kaloper:2014dqa} (see also the review~\cite{Padilla:2015aaa}). 
In that context, the effective field theory expansion has to have 
another global scale, $s$, 
so that the sequestering action is proportional to 
\begin{equation}
 \int_X \sqrt{-g} \Big[\frac{R}{2G} + s^4 L_m (s^{-2} g^{ab}) + \frac{\Lambda}{G}\Big]
 + \sigma\Big(\frac{\Lambda}{s^4 \mu^4}\Big),
\end{equation}
where $\mu$ is a mass scale and $\sigma(\frac{\Lambda}{s^4 \mu^4})$ is a
global interaction that is not integrated over
\cite{Kaloper:2014dqa, Padilla:2015aaa}. This can be provided by our set up: Start with bilocal
fields  $\phi(x, \tx)$~\cite{Freidel:2016pls}, and replace the dual labels $\tx$ and also $\lambda$ (in a coarsest approximation) by the global dynamical scale $s \sim \Delta {\tx} \,{\sim}\,\l^2 \Delta {x}^{-1} $.
Also, normal ordering produces $\sigma$.
This is 
an effective realization of the sequester mechanism 
in a  non-commutative phase of string theory. 
Furthermore, the intrinsic non-commutativity of the zero modes $x$ and $\tx$~\eqref{e:CnCR1} corrects  the zeroth order results in $\lambda$ in several ways. 
In particular, it is natural to ask whether the non-zero $\omega^{AB}$ in~\eqref{e:CnCR}  stabilizes the cosmological constant directly on the level of the effective non-commutative action.
 The fully non-commutative analysis is intricate, but 
for conformally flat metrics, $g_{\mu \nu} = \phi^2 \eta_{\mu \nu}$, the action~\eqref{e:ncEH}--\eqref{e:TsSd} produces a non-commutative $\L \phi^4$ theory, which is a natural non-commutative generalization of 
the effective action for conformal metrics  $\int_X (\partial_{\mu} \phi\,\partial^{\mu} \phi + \frac{\L}{3}  \phi^4)$, with the
non-commutative product depending on $\lambda$.
 Unlike the commutative limit of the theory, the beautiful results of Grosse and Wulkenhaar~\cite{Grosse:2012uv} demonstrate the non-perturbative solvability of the above non-commutative $\L \phi^4$ theory, explicitly showing the finite renormalization of $\L$ in terms of the bare coupling.
 At least in this highly simplified, conformal degree limit, non-commutativity thus can afford a small, radiatively and perhaps even non-perturbatively stable cosmological constant for the non-commutative form of the ``doubled'' effective action.
The non-commutativity scale $\l$ is naturally related to the Planck scale; see~\eqref{e:CnCR1}. 

However, non-commutative field theories have both UV and IR scales and 
the effective description is defined by expanding around self-dual fixed points, 
and 
it is organized by keeping  both 
the Wilsonian UV cutoff as well as the IR scale. This clearly meshes nicely with
the UV and IR aspects of the see-saw formula.
Identifying $M_\L$ and $M_P$ as the IR and UV cut-offs, respectively, the double-scale RG flow 
identifies a self-dual fixed point~\cite{Douglas:2001ba, Grosse:2004yu}. 
Given that the phase-space formulation~\cite{Freidel:2013zga, Freidel:2015uug, Freidel:2016pls, 
 Freidel:2017xsi, Freidel:2017wst} is a T-duality covariant description of string theory, this naturally relates $M_P\,{\to}\,M^2/M_P$ under T-duality.
The prediction of our models~\cite{rBHM6, rBHM7} $M_\L\,{\sim}\,M^2/M_P$ then satisfies these conditions, with $M_P\,{\sim}\,\e\,{=}\,1/{\lambda}$ the fundamental energy scale corresponding to the fundamental length $\l$, 
which is consistent with observations 
provided $M$ is a \TeV\ scale. 
We emphasize that the usual spacetime discussion 
of string theory is compatible with local effective field theory,
which does not account for the radiative stability of vacuum energy. What we argue is that
this feature of string theory is an artifact of a spacetime description, which is not generic.
The generic formulation of string theory is doubled and generalized-geometric, and intrinsically
non-commutative, and it leads to an effective field theory that is sequestered, and thus, to leading order,
to a radiatively stable vacuum energy. (Including further corrections due to intrinsic
non-commutativity.) Only in a singular limit in which one neglects the intrinsic non-commutativity
and works only within a spacetime section of the general doubled description does one find the usual effective 
field theory, with a spacetime interpretation, and the usual questions regarding the existence of
de~Sitter background in string theory
\cite{Danielsson:2018ztv}.

\paragraph{Outlook:} In the context of F-theory (see also~\cite{Heckman:2018mxl}), in which our cosmic-string-like solution naturally finds its habitat, we are 
led to contemplate a non-commutative phase of F-theory (see also~\cite{Heckman:2010pv}, and~\cite{Berenstein:2000jh}) 
in which this approach to dark energy in
string theory is realized.
Our effective string-like~\cite{Polchinski:1991ax} construction implies that the breaking of supersymmetry (which underlies the canonical commutative formulation of F-theory) is
related, in our stringy-cosmic-string-like solution, to the emergent de~Sitter solution, in the full non-commutative (and non-supersymmetric) phase of F-theory\footnote{The overall physics here may be linked to the old observation of Witten~\cite{Witten:1994cga}, that supercharges need not be globally defined in the presence of conical defects, and the mass splitting between superpartners is controlled by the strength of the conical defect. For the corresponding four-dimensional generalization and relation to the cosmological constant, see~\cite{Jejjala:2002we}.}.
When discussing a non-commutative phase of F-theory it is natural to invoke the IIB matrix model~\cite{Ishibashi:1996xs}, which describes $N$ D-instantons (and is by T-duality related to the Matrix model of M-theory~\cite{Banks:1996vh}).
Given~\eqref{e:MSA}, we can suggest a {\it new covariant} non-commutative matrix model formulation of F-theory, by writing in the large $N$ limit $\pa_\s\X^C = [\X, \X^C]$ 
(and similarly for $\pa_{\tau}\X^B$) in terms of commutators of two (one for $\pa_\s\X^C$ and one for $\pa_{\tau}\X^C$) extra $N\,{\times}\,N$ matrix valued chiral $\X$'s:
\begin{equation}
 S_{\text{ncF}}\,{=}\,\frac{1}{4\pi} \mathrm{Tr} 
[{\X}^{a},  {\X}^{b}] [{\X}^{c},  {\X}^{d}]  f_{abcd},
\end{equation}
where instead of $26$ bosonic $\X$ matrices one would have $28$, with supersymmetry emerging in 10($+$2) dimensions from this underlying
 bosonic formulation.
By T-duality, the new covariant M-theory matrix model reads as 
\begin{equation}
 S_{\text{ncM}}\,{=}\,\frac{1}{4\pi}  
\int_{\tau} \mathrm{Tr} \big(\pa_{\tau} \X^i [{\X}^{j},  {\X}^{k}] g_{ijk} - [{\X}^{i},  {\X}^{j}][{\X}^{k},  {\X}^{l}] h_{ijkl}\big),
\end{equation}
with 27 bosonic $\X$ matrices, with supersymmetry emerging in 11 dimensions. Note that the information about
 $\w_{AB},\eta_{AB}$ and $H_{AB}$
 combine into the new dynamical backgrounds $g_{ijk}$ and $h_{ijkl}$ of M-theory,
 and fully unify into $f_{abcd}$ of F-theory.

\paragraph{Conclusion:}
We have argued that the generic
formulation of string theory leads naturally to dark energy, represented by a positive cosmological constant {\it to lowest order} and
the intrinsic stringy non-commutativity is the new crucial ingredient responsible for its radiative stability.
Also, an effective cosmic-string-like solution of F-theory naturally fits into this formalism
and it leads to a see-saw-like formula for the cosmological constant.
Note also that Starobinsky inflation~\cite{Starobinsky:1980te} may appear as a natural product of the higher order terms in the $\lambda$ expansion that,
after integrating over $\tilde{x}$ can result  in $\int_X \sqrt{-g} (R + a R^2)$, at the next to leading order in $\lambda$. Starobinsky inflation 
beautifully fits the observed data~\cite{Akrami:2018odb}, and is non-supersymmetric ---
which is consistent with the supersymmetry-breaking nature of our construction.
Finally, our discussion naturally relates to
the observationally supported proposal for dark matter quanta that are sensitive to dark energy~\cite{Ho:2010ca}.

\paragraph{Acknowledgements:}
PB 
thanks the Simons Center for Geometry
  and Physics and the CERN Theory Group, 
 and TH 
the Department of Physics, University of Maryland, 
 and the Physics Department of 
 the University of Novi Sad, Serbia, for hospitality and resources. DM thanks L. Freidel and R. G. Leigh for collaboration
and the Julian Schwinger Foundation for support.

%
%
\begingroup
\frenchspacing\raggedright
\endgroup

\end{document}